\documentclass[twoside,slac_one]{revtex4}
\usepackage{graphicx}
\usepackage{fancyhdr}
\usepackage{amsmath} 
\usepackage{bm}
\usepackage{amsxtra}
\usepackage{amssymb}
\usepackage{amsthm}
\usepackage{latexsym}
\usepackage{lscape}

\usepackage[percent]{overpic}
\usepackage{subfigure}
\RequirePackage{xspace}
\usepackage{relsize}

\def\pep2{\mbox{PEP-II} \xspace}
\def\BF{$B$ Factory}
\def\abf {asymmetric \BF}

\def\babar{\mbox{\slshape B\kern-0.1em{\smaller A}\kern-0.1em
  B\kern-0.1em{\smaller A\kern-0.2em R}}\xspace}

\def\invfb  {\ensuremath{\mbox{\,fb}^{-1}}\xspace}
\def\invpb  {\ensuremath{\mbox{\,pb}^{-1}}\xspace}

\def\qqbar {\ensuremath{q\overline q}\xspace}

\def \Bl{\mathcal{B}(B_s \rightarrow \ell \nu  X)}

\def\Y#1S{\ensuremath{\Upsilon{(#1S)}}\xspace}

\def\Bbar  {\kern 0.18em\overline{\kern -0.18em B}{}\xspace}

\def\BB   {\ensuremath{B\Bbar}\xspace} 

\def\Kbar {\kern 0.2em\overline{\kern -0.2em K}{}\xspace}

\def\Kz  {\ensuremath{K^0}\xspace}
\def\Kzb  {\ensuremath{\Kbar^0}\xspace}
\def\KzKzb {\ensuremath{\Kz \kern -0.16em \Kzb}\xspace}

\newcommand{\gev}{\ensuremath{{\mathrm{\,Ge\kern -0.1em V}}}\xspace}
\newcommand{\mev}{\ensuremath{{\mathrm{\,Me\kern -0.1em V}}}\xspace}

\newcommand{\gevc}{\ensuremath{{\mathrm{\,Ge\kern -0.1em V\!/}c}}\xspace}
\newcommand{\mevc}{\ensuremath{{\mathrm{\,Me\kern -0.1em V\!/}c}}\xspace}
\newcommand{\gevcc}{\ensuremath{{\mathrm{\,Ge\kern -0.1em V\!/}c^2}}\xspace}
\newcommand{\mevcc}{\ensuremath{{\mathrm{\,Me\kern -0.1em V\!/}c^2}}\xspace}

\newcommand{\stat}{\ensuremath{\mathrm{(stat)}}\xspace}
\newcommand{\syst}{\ensuremath{\mathrm{(syst)}}\xspace}

\begin{document}

\title{Recent Results in Semileptonic $B$ Decays With \babar}

%

\author{B. Hamilton}
\affiliation{Department of Physics, University of Maryland, College Park, MD, USA}

\begin{abstract}
In this note, recent results of studies of semileptonic B meson decays from \babar\ are discussed and preliminary results given. In particular, a recent measurement of $\mathcal{B}(B \to D^{(*)}\tau \nu)$ and the ratio $\mathcal{B}(B \to D^{(*)}\tau \nu)/\mathcal{B}(B \to D^{(*)}\ell \nu)$ is presented. For the $D^*$ mode, a branching fraction of $1.79\pm0.13\stat\pm0.17\syst$ is found, with a ratio of $0.325\pm0.023\stat\pm0.027\syst$. For the $D$ mode, the results are $1.04\pm0.12\stat\pm0.14\syst$ and $0.456\pm0.053\stat\pm0.056\syst$, respectively.  In addition, a study of $B_s$ production and semileptonic decays using data collected in a center-of-mass energy region above the \Y4S resonance is discussed. The semileptonic branching fraction $\mathcal{B}(B_s \to \ell \nu X)$ is measured to be $9.9{ }^{+2.6}_{-2.1}\stat{ }^{+1.3}_{-2.0}\syst$. 
\end{abstract}

\maketitle

\thispagestyle{fancy}


\section{Introduction to the \babar\ Experiment}

The \babar\ experiment at SLAC's \pep2 \abf\ ran from 2000-2008 measuring the products of asymmetric $e^+ e^-$
 collisions. During this time, it gathered a dataset consisting of 432.89\invfb at a center-of-mass (CM) energy equal to the \Y4S mass.
 In addition, data were taken at the 
 \Y3S and \Y2S resonances, as well as CM energy points below each resonance. Finally, in 2008 a scan was performed covering
 the center-of-mass energy ($E_{\rm CM}$) region from 10.54\gev to 11.20\gev in 5\mev steps, followed by additional
 running at points spread around the $\Upsilon(11020)$ peak.

A detailed description of the \babar\ detector can be found in Ref. \cite{Aubert:2001tu}. Tracking of charged particles 
is provided by the combination of an inner, five-layer silicon tracker (SVT) and an outer, 40-layer drift chamber 
(DCH). Momentum information is provided by the use of a 1.5-Tesla solenoidal magnetic field surrounding the 
tracking subdetectors. Photon and electron energy measurements are provided by a CsI(Tl) electromagnetic 
calorimeter. Charged particle identification at \babar\ is achieved using a likelihood-based approach combining 
specific ionization ($dE/dx$) measurements from the DCH and SVT, energy deposition in a $CsI(Tl)$ electromagnetic 
calorimeter (EMC), and measurements of Cherenkov radiation produced in the detector of 
internally-reflected Cherenkov light (DIRC). An instrumented flux return (IFR) provides additional information for
discriminating $\mu$ from $\pi$.

\section{Measurements of $B \to D^{(*)}\tau\nu$}
\subsection{Introduction}
Semileptonic decays of $B$ mesons provide a valuable probe of the electroweak sector of the standard model. 
Lepton universality in the standard model means that differences in the semileptonic branching fractions to $e$, 
$\mu$, and $\tau$ arise only through the mass splitting among the three. Therefore, deviations from the predicted 
ratios of branching fractions would be a sign of new physics. However, due to their more complicated nature, 
$B$ semileptonic decays to $\tau$ are not as precisely measured as those to $e$ or $\mu$. 
The first study presented here is an update of the \babar\ measurement
of exclusive $B$ decays to $D^{*}\tau\nu$, with improvements in both the size of the dataset used and the efficiency 
of the selection relative to the previous 
\babar\ measurement \cite{oldresult}.

The variable used to study these semileptonic decays to $\tau$ is the ratio of the semileptonic branching fraction to $\tau$ to the branching fraction to the light leptons $\ell = e \text{ or } \mu$
\begin{equation}
R(D^{(*)}) \equiv \frac{\mathcal{B}(B \to D^{(*)}\tau \nu)}{\mathcal{B}(B \to D^{(*)}\ell \nu)},
\end{equation}
which is both theoretically well-controlled and allows for certain experimental systematic uncertainties to cancel in the ratio. Standard model predictions predictions for this quantity are given in Ref. \cite{RD} as $R(D) = 0.31\pm0.02$ and $R(D^*)=0.25\pm0.02$. This quantity is potentially quite sensitive to charged currents induced by new physics, for example the existence of an extra Higgs doublet \cite{tanaka}.

\subsection{Event Reconstruction}
For this measurement, events with a single well-identified $e$ or $\mu$ candidate are selected. The rest of the tracks in the event are used to reconstruct a $B$ and a $D^(*)$ candidate. The $B$ candidate, referred to as the ``tag" $B$, is reconstructed in the event from one of over 1000 possible hadronic decay modes, constraining all the daughter tracks to come from the event's primary vertex. For each $B_{\rm tag}$ candidate, the following quantities are used to suppress combinatoric backgrounds: the difference of the reconstructed $B$ energy and the beam energy in the center of mass, and the mass of the $B$ candidate calculated using the known collision energy
\begin{equation}
\Delta E = E-E^*_{\rm beam}
\qquad
m_{\rm ES} = \sqrt{{E^{2}_{\rm CM}}/4-\vec{p}^2}.
\end{equation} For this measurement, $B_{\rm tag}$ candidates are required to satisfy $|\Delta E| < 0.072$ and $m_{\rm ES} > 5.27\gev$. 

The remaining tracks in the event are used to form a $D$ candidate decaying via $D\to K K^+$, $D \to K \pi$, $D \to K \pi\pi$ or $D \to K \pi\pi\pi$ where $K=K^0_s$ or $K^-$. If possible, $D^*$ candidates are reconstructed via $D^{*+} \to D^+\pi^0$, $D^{*+} \to D^0\pi^+$, $D^{*0} \to D^0\pi^0$ or $D^{*+} \to D^0\gamma$. Both $D$ and $D^*$ reconstructions involve updating the fit of their daughter tracks using the added constraint that their masses be equal to the world average values \cite{PDG}. This improves the resolution on the missing energy and momentum in the event.

In cases where an event may be reconstructed in more than one way, one of the possibilities is chosen by the use of extra energy deposited in the calorimeter. The extra energy in an event is defined as scalar sum of the energy deposited in the EMC that is both isolated from any charged tracks and not used as a photon candidate in the reconstruction of the $B$ or $D^*$. The reconstruction which minimizes this quantity is selected, while the others are discarded.

The final part of the reconstruction is the construction of a $D^{**}$ control sample. Unused photon candidates in the EMC are combined to form $\pi^0$ candidates. Valid $\pi^0$ candidates are required to have an invariant mass between 120\mev and 150\mev. Events with these $\pi^0$ candidates are used to estimate the background coming from $B$ decays to highly excited charm states, which subsequently feed-down into the $D$ or $D^*$ samples due to errors in reconstruction and limited detector acceptance.

Semileptonic decays to $\tau$ are distinguished by examining the kinematics of the lepton candidate and the missing energy and momentum in the event. The true value of the variable $q^2 \equiv (E_{\rm miss} + E_\ell)^2 + (\vec{p}_{\rm miss} + \vec{p}_\ell)^2$ is bounded from below by $m_\ell$, so requiring $q^2 > 4 \gev^2 > m^2_\tau$ removes much of the background from semileptonic decays to $e$ and $\mu$.

Backgrounds in this measurement are further controlled by the use of four multivariate selectors, trained to discriminate signal events against $\BB$, continuum, and charge-crossfeed backgrounds in the four signal samples.  Eight additional selectors are used to remove background in the four $D^{(*)}\pi^0\ell\nu$ samples, with one selectors for each channel trained against background from other semileptonic decays and one trained against $\BB$ combinatoric background.

\subsection{Fit}

The result is extracted using a two-dimensional fit in the variables $p^*_\ell$ and $m^2_{\rm miss}$, respectively the momentum of the lepton in the center-of-mass frame and the invariant mass of the missing energy and momentum in the event. The fit is performed simultaneously on eight data samples: four signal $D^{(*)}\tau\nu$ samples and four $D^{(*)}\pi^0\ell\nu$ samples. Signal and background shapes are taken from simulation using non-parametric kernel estimators \cite{kernel}. The relative normalization of background from charge-crossfeed and continuum events are fixed using simulation in all samples. The \BB background is fixed in the signal and normalization samples, but not the $D^{(*)}\pi^0\ell\nu$ samples.

Simulation is used to create constraints between the various samples, for example the yield of $B\to D^*\tau\nu$ events determines the amount of $B\to D^*\tau\nu$ feed-down into the $D$ sample.  These constraints are used for all such yields except for the feed-down of $B\to D^*\ell\nu$ into $D$, which is left floating due to the very high statistics in this channel. This leaves 22 free parameters in the fit: the primary yield in each sample (i.e. $B\to D\tau\nu$ or $B\to D\ell\nu$ in the $D$ sample) (twelve parameters), the \BB background in the four $D^{(*)}\pi^0\ell\nu$ samples (four parameters), and the two $B\to D^*\ell\nu$ feed-down yields already mentioned.

\begin{figure*}[ht]
\centering
\begin{overpic}[width=135mm]{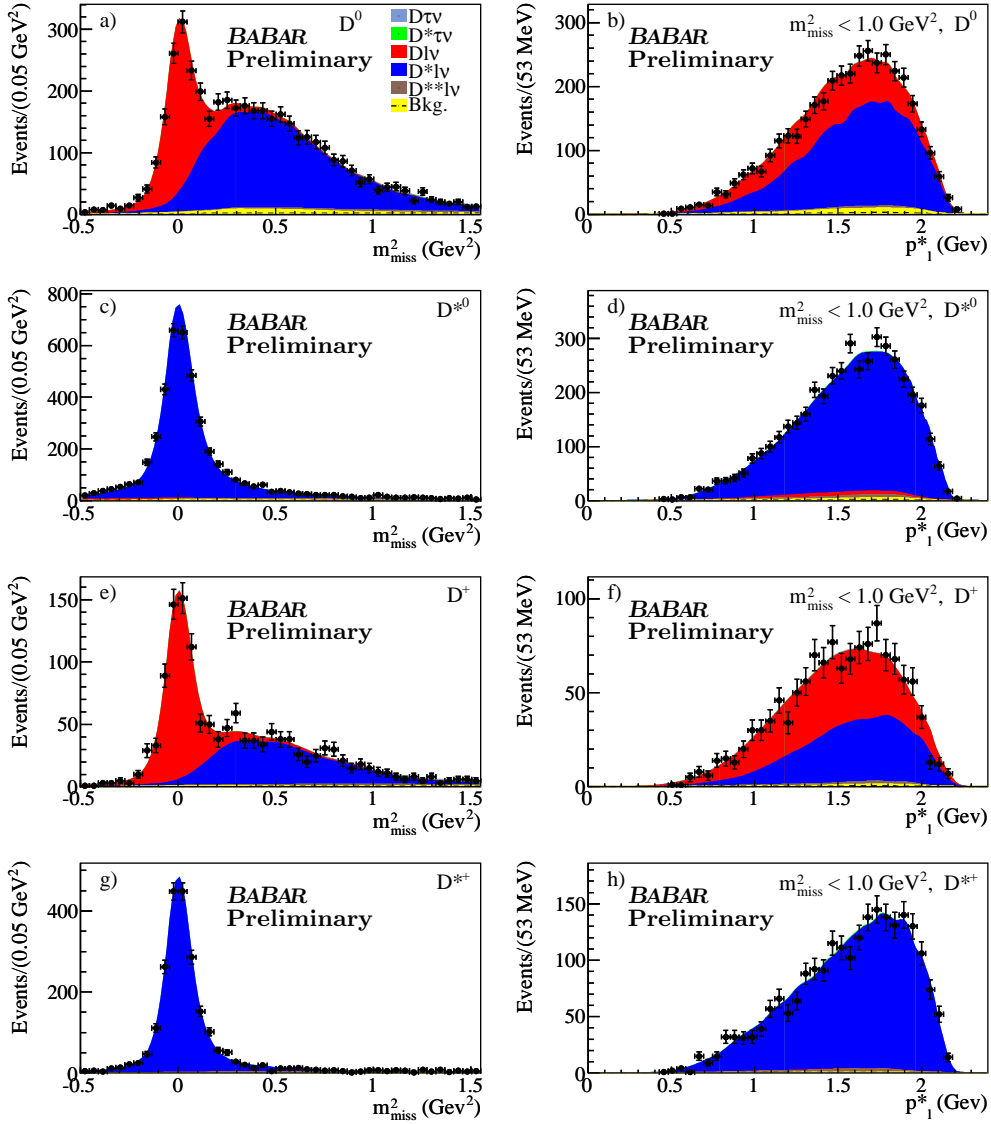}
\put(20,95){\bf \babar}\put(20,93){\bf Preliminary}
\put(55,95){\bf \babar}\put(55,93){\bf Preliminary}
\put(20,70){\bf \babar}\put(20,68){\bf Preliminary}
\put(55,70){\bf \babar}\put(55,68){\bf Preliminary}
\put(20,45){\bf \babar}\put(20,43){\bf Preliminary}
\put(55,45){\bf \babar}\put(55,43){\bf Preliminary}
\put(20,20){\bf \babar}\put(20,18){\bf Preliminary}
\put(55,20){\bf \babar}\put(55,18){\bf Preliminary}
\end{overpic}
\caption{Fit in the normalization region} \label{fig:norm}
\end{figure*}

\begin{figure*}[ht]
\centering
\begin{overpic}[width=135mm]{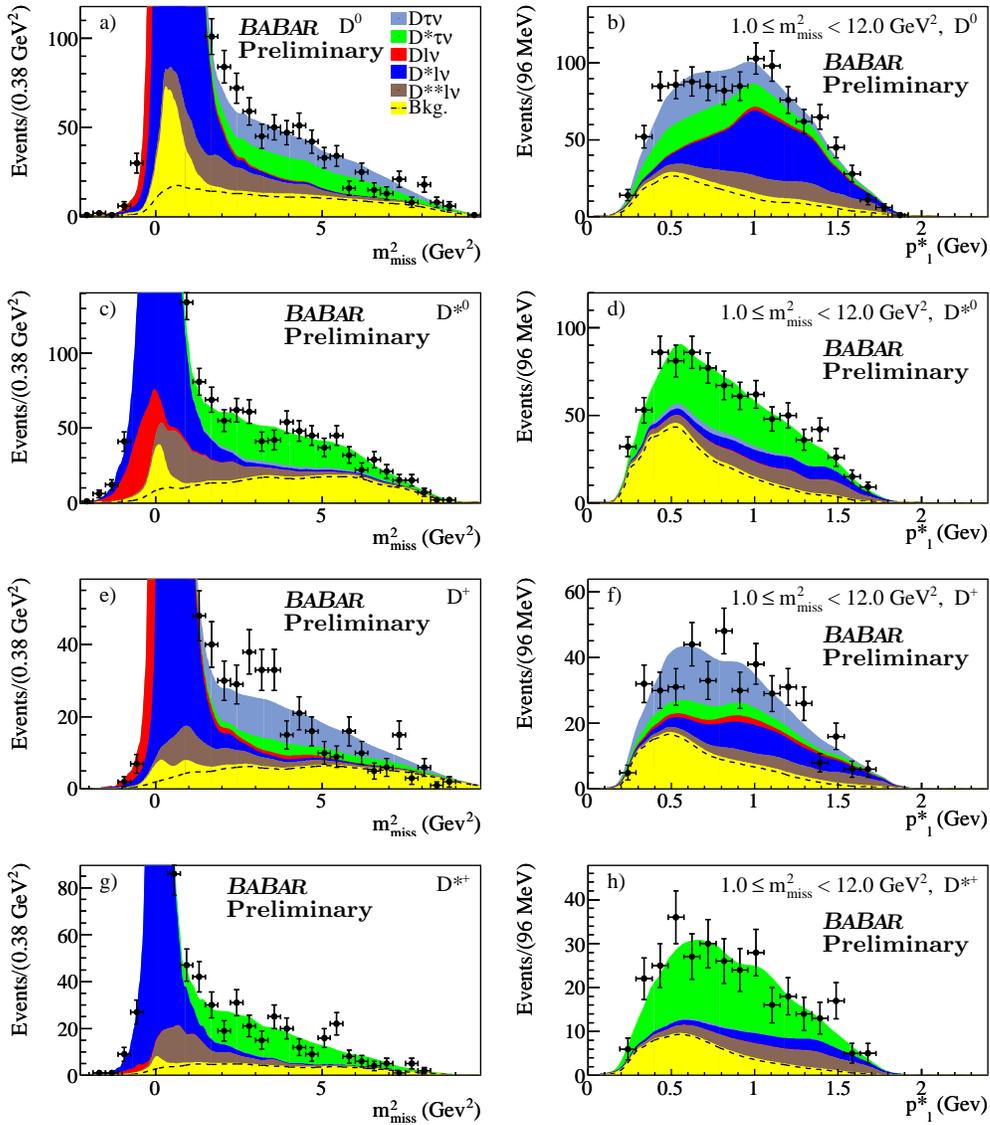}
\put(21,96){\bf \babar}\put(21,94){\bf Preliminary}
\put(72,93){\bf \babar}\put(72,91){\bf Preliminary}
\put(25,71){\bf \babar}\put(25,69){\bf Preliminary}
\put(72,68){\bf \babar}\put(72,66){\bf Preliminary}
\put(25,46){\bf \babar}\put(25,44){\bf Preliminary}
\put(72,43){\bf \babar}\put(72,41){\bf Preliminary}
\put(20,21){\bf \babar}\put(20,19){\bf Preliminary}
\put(72,18){\bf \babar}\put(72,16){\bf Preliminary}
\end{overpic}
\caption{Fit in the signal region} \label{fig:sig}
\end{figure*}

In addition to the 22 parameter fit (referred to as the {\it unconstrained fit}), a separate fit is performed in which the yields of the charged and neutral $D^(*)$ components are constrained using known isospin relations in the decays of charged and neutral $B$ mesons. This {\it isospin constrained fit} has 13 free parameters. The unconstrained fit is shown with the normalization region projected in Fig. \ref{fig:norm} and the signal region in \ref{fig:sig}.

\subsection{Results}
The preliminary results are summarized in table \ref{tab:tauresult}. Both the $D$ and $D^*$ are consistent with the standard model expectation at the level of 1.8 standard deviations. This study represents the first measurement of $\mathcal{B}(B\to D\tau \nu)$ at a precision above 5$\sigma$.

\begin{table*}[t]
\begin{center}
\caption{Preliminary results of $D^{*}\tau\nu$ measurement.}
\begin{tabular}{| c |c|c|c|c|c|c|} \hline
 & \multicolumn{3}{c|}{$\mathbf{D^*}$} & 
\multicolumn{3}{c|}{$\mathbf{D}$} \\\hline
\textbf{Charge State} & \textbf{Yield} & $\mathcal{B}(B\to D^*\tau\nu)$ (\%) & $R(D^*)$ & \textbf{Yield} & $\mathcal{B}(B\to D\tau\nu)$  (\%)& R(D) \\\hline
\textbf{Neutral} & $511\pm 48$ & $1.79\pm0.17\pm0.14$ & $0.341\pm0.030\pm0.028$ & $226\pm39$ & $0.96\pm0.16\pm0.14$ & $0.422\pm0.074\pm0.059$ \\\hline
\textbf{Charged} & $220\pm23$ & $1.82\pm0.19 \pm 0.17$ & $0.356\pm0.038\pm0.032$ & $139\pm21$ & $1.08\pm0.19\pm 0.15$ & $0.513\pm0.081\pm0.067$ \\\hline
\textbf{Constrained} & $730\pm50$ & $1.79\pm0.13\pm0.17$ & $0.325\pm0.023\pm0.027$ & $368\pm42$ & $1.04\pm0.12\pm0.14$& $0.456\pm0.053\pm0.056$\\\hline
\end{tabular}
\label{tab:tauresult}
\end{center}
\end{table*}

\clearpage
\section{Measurements of $B_s$ Production and Decay}
\subsection{Introduction}
The study of $B_s$ mesons can provide another opportunity to study semileptonic $b$ quark decays, 
though it is more difficult to study them in the clean environment of an $e^+e^-$ collider due to the lack of an efficient production mechanism similar to $e^+e^- \to \Y4S \to \BB$. As a result, information on inclusive $B_s$ decays remains scarce \cite{PDG}. The second study presented here uses data taken by \babar\ with the \pep2 \abf scanning the center-of-mass (CM) energy region above the $\Y4S$. This is used to measure $B_s$ production in this region, parameterized by the ratio
\begin{equation}
f_s \equiv \frac{N_{B_s}}{N_{B_s}+N_{B_d}+N_{B_u}},
\end{equation}
and the inclusive semileptonic branching fraction $\mathcal{B}(B_s \to \ell\nu X)$. The scan consists nominally of 25\invpb points in 5\mev steps from 10.54\gev to 11.2\gev, plus six additional 100\invpb points in the region of the $\Upsilon(11020)$. This represents a total of about 3.15\invfb above the $B_s\Bbar_s$ threshold.

In order to distinguish \BB events from $B_s\Bbar_s$ events, the large difference in $\phi$ meson production between $B_s$ and $B$ decays is used. This difference arises due to the large $D_s \to \phi$ branching fraction compared to $D \to \phi$ along with the large $D_s$ multiplicity in $B_s$ events. The expected contribution of the favored spectator decay chain $B_s \to D_s \to \phi$ is on the order of 15\%, compared to the measured inclusive branching fraction $\mathcal{B}(B \to \phi X) = 3.43\%$ \cite{PDG}. This difference can be exploited so that $B_s$ and $B$ contributions can be separated in an inclusive measurement without a priori knowledge of the relative production rate of $B_s$ vs $B$.

\subsection{Measurements}

For this measurement, the scan data are grouped into 15\mev wide $E_{\rm CM}$ bins. In each bin, the following quantities are measured, normalized to the number of  $e^+e^- \to \mu^+\mu^-$ events:
\begin{itemize}
\item The number of events passing a multi-hadronic event selection designed to reject continuum $e^+e^- \to \qqbar,\; q=u,d,s,c$ events.
\item The number of events passing the multi-hadronic selection which contain a $\phi$ meson.
\item The number of events passing the multi-hadronic selection and containing a $\phi$ meson which also contain a well-identified $e$ or $\mu$ candidate with momentum greater than 0.9\gev in the CM frame.
\end{itemize}
These are here referred to as the event yield, the $\phi$ yield and the $\phi$-lepton yield, respectively.

\begin{figure}[tbh]
\centering
	\subfigure{\begin{overpic}[width=80mm]{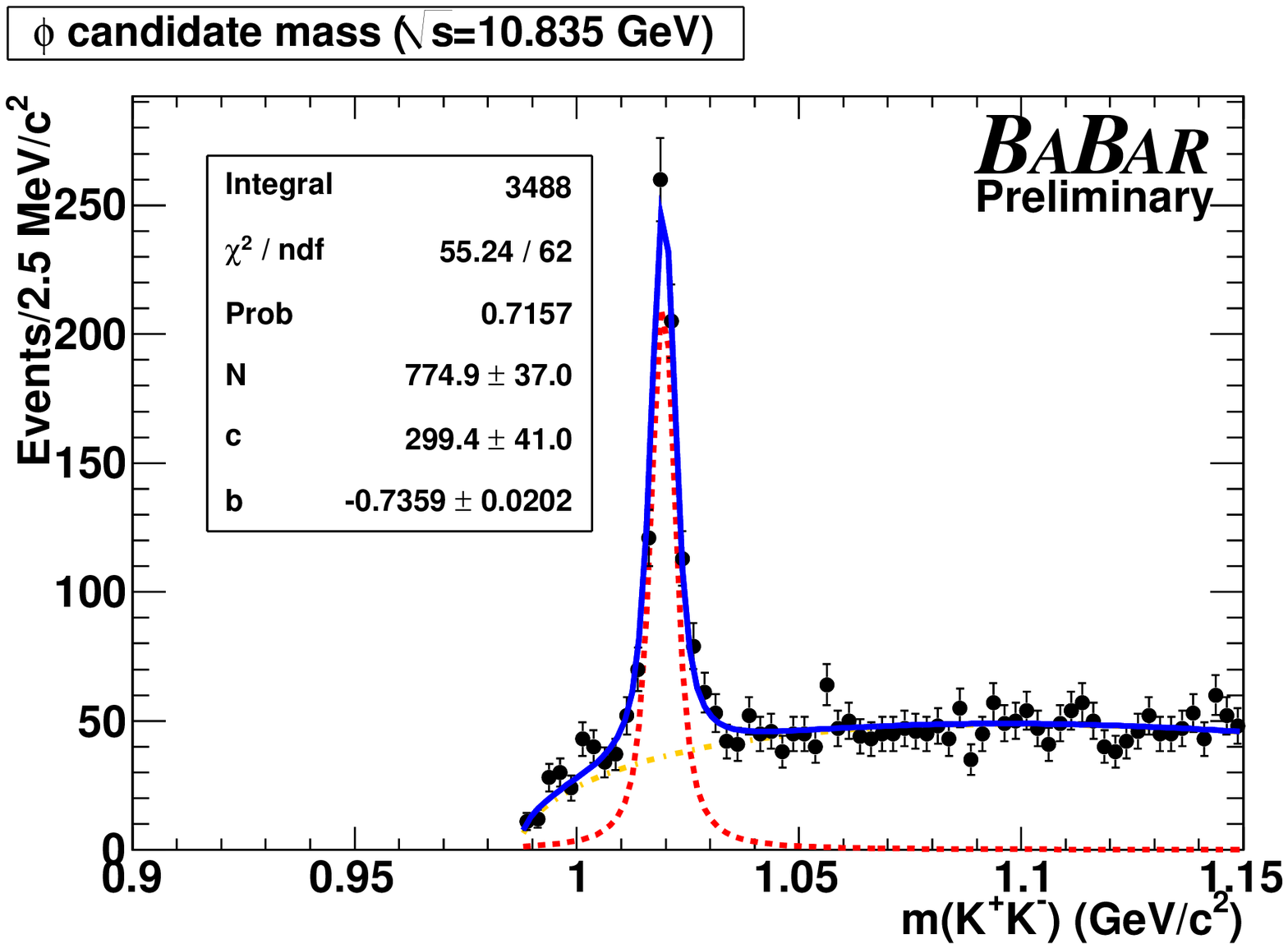}\put(50,55){(a)}\end{overpic}}\\
	\subfigure{\begin{overpic}[width=80mm]{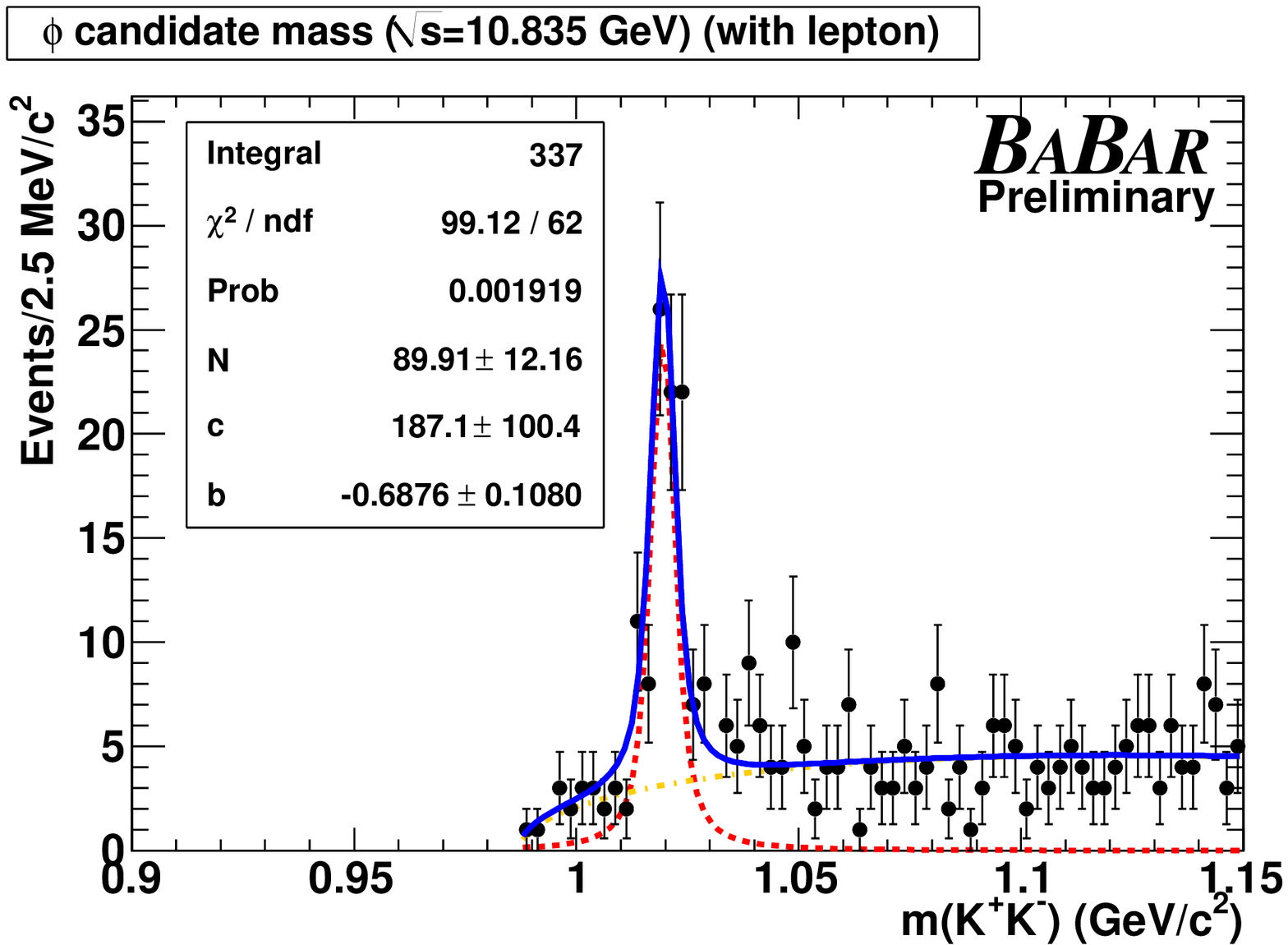}\put(50,55){(b)}\end{overpic}}
\caption[s]{ Invariant mass distribution of (a) $\phi \rightarrow K^+K^-$ candidates in  $E_{\rm CM}$ bin (10.8275\gev $\leq E_{\rm CM} \leq$ 10.8425\gev ) and 
(b) invariant mass distribution of $\phi \rightarrow K^+K^-$ candidates in event containing a high momentum lepton candidate in the same $E_{\rm CM}$ range. The background
shape is shown by the broken orange curve, the peak by the broken red curve, and the total fit by the unbroken blue curve.}
\label{fig:phifits}
\end{figure}

Candidate $\phi$ mesons are first formed by combining pairs of oppositely-charged $K$ candidate tracks. In each event, the $\phi$ candidate which has the best identified daughter $K$ candidates is selected. The invariant mass distribution of these $\phi$ candidates is fit to a signal plus background distribution to extract the number of candidates in the $\phi$ mass peak for a particular $E_{\rm CM}$ bin. In particular, the function
\begin{equation}
f(m_{KK};N,b,c) = NV(m_{KK};m_\phi,\Gamma_\phi,\sigma) + Nc(1+b m_{KK}) \sqrt{1+ \left( \frac{2m_K}{m_{KK}} \right)^2}\label{eqn:fit}
\end{equation}
is used, where $V(m_{KK};m_\phi,\Gamma_\phi,\sigma)$ is a Voigt profile (convolution of a Gaussian resolution function with a Breit-Wigner peak) with central value and Breit-Wigner width equal to the world average $\phi$ mass  and width $m_\phi$ and $\Gamma_\phi$ in Ref. \cite{PDG}. The gaussian resolution $\sigma$ is fixed to the average value across the scan. The background function used is a simple linear factor parameterized by $b$ times a threshold function at twice the $K$ mass $m_K$, with relative normalization $c$. Examples of this fit are shown in Fig. \ref{fig:phifits}.

\begin{figure}[bth]
\centering
\includegraphics[width=80mm]{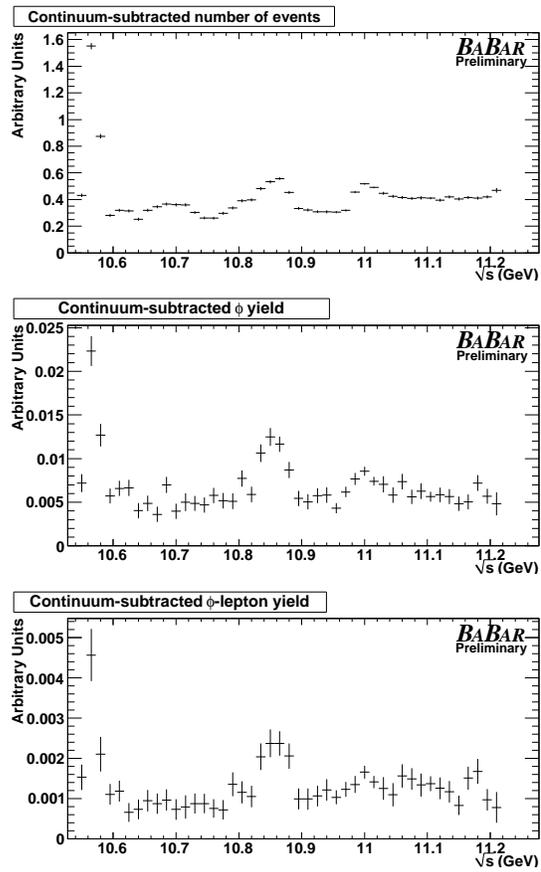}
\caption{Event, $\phi$ and $\phi$-lepton yields after continuum subtraction. Note that efficiencies have not be removed at this point, so the overall normalization is arbitrary.}
\label{fig:yields}
\end{figure}

The three yields are also measured in a sample of data taken by \babar\ below the \BB threshold in order to determine the contribution from continuum $e^+e^- \to \qqbar,\; q=u,d,s,c$ events. The yields in the continuum data are corrected for running efficiencies point-by-point and subtracted from each bin, leaving the contributions from \BB and $B_s\Bbar_s$, shown in Fig. \ref{fig:yields}. The \BB and $B_s\Bbar_s$ contributions to the yields may be written schematically as
\begin{align}
C_{\rm MH} \!={ }& R_B\left[f_s \epsilon^s_{\rm MH} + (1-f_s) \epsilon_{\rm MH}\right]\label{eqn:c1}\\
C_{\phi} \;\;\,={ }& R_B\left[f_s \epsilon^s_{\phi} P(B_s \Bbar_s \to \phi X) + (1-f_s) \epsilon_{\phi} P(\BB \to \phi X) \right]\label{eqn:c2}\\
C_{\phi\ell} \;\,={ }& R_B\left[f_s \epsilon^s_{\phi\ell} P(B_s \Bbar_s \to \phi \ell \nu X) + (1-f_s) \epsilon_{\phi\ell} P(\BB \to \phi \ell \nu X) \right]\label{eqn:c3}
\end{align}
with $R_B \equiv \sigma(e^+e^- \to B_q\Bbar_q,\; q=u,d,s)/\sigma(e^+e^- \to \mu^+\mu^-)$, and $\epsilon^{(s)}_X$ the efficiency for $B_{(s)}$ events to contribute to the event, $\phi$ or $\phi$-lepton yield. The contributions from \BB events, $\epsilon_{\phi} P(\BB \to \phi X)$ and $\epsilon_{\phi\ell} P(\BB \to \phi \ell \nu X)$ are determined in \babar's \Y4S dataset, with efficiencies corrected for energy dependence using simulation. The quantity $P(B_s \Bbar_s \to \phi X)$ is estimated using known branching fractions from \cite{PDG} along with an estimate of direct $B_s \to \phi$ transitions. $P(B_s \Bbar_s \to \phi X)$ is similarly estimated, but left as a function of the unknown branching fraction to be measured, $\Bl$.

\subsection{Extraction of Results}

\begin{figure}[hbt]
\centering
\includegraphics[width=80mm]{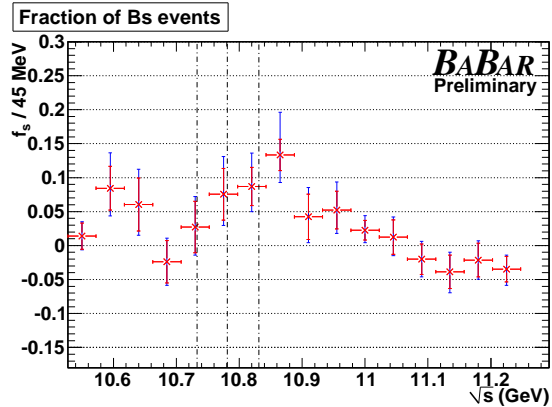}
\caption{Results for the fraction $f_s$ binned in 45 \mev steps. The inner (red) error bars show the statistical uncertainty, while the outer (blue) error bars show the statistical and systematic uncertainties added in quadrature.}
\label{fig:fs}
\end{figure}

\begin{figure}[hbt]
\centering
\includegraphics[width=80mm]{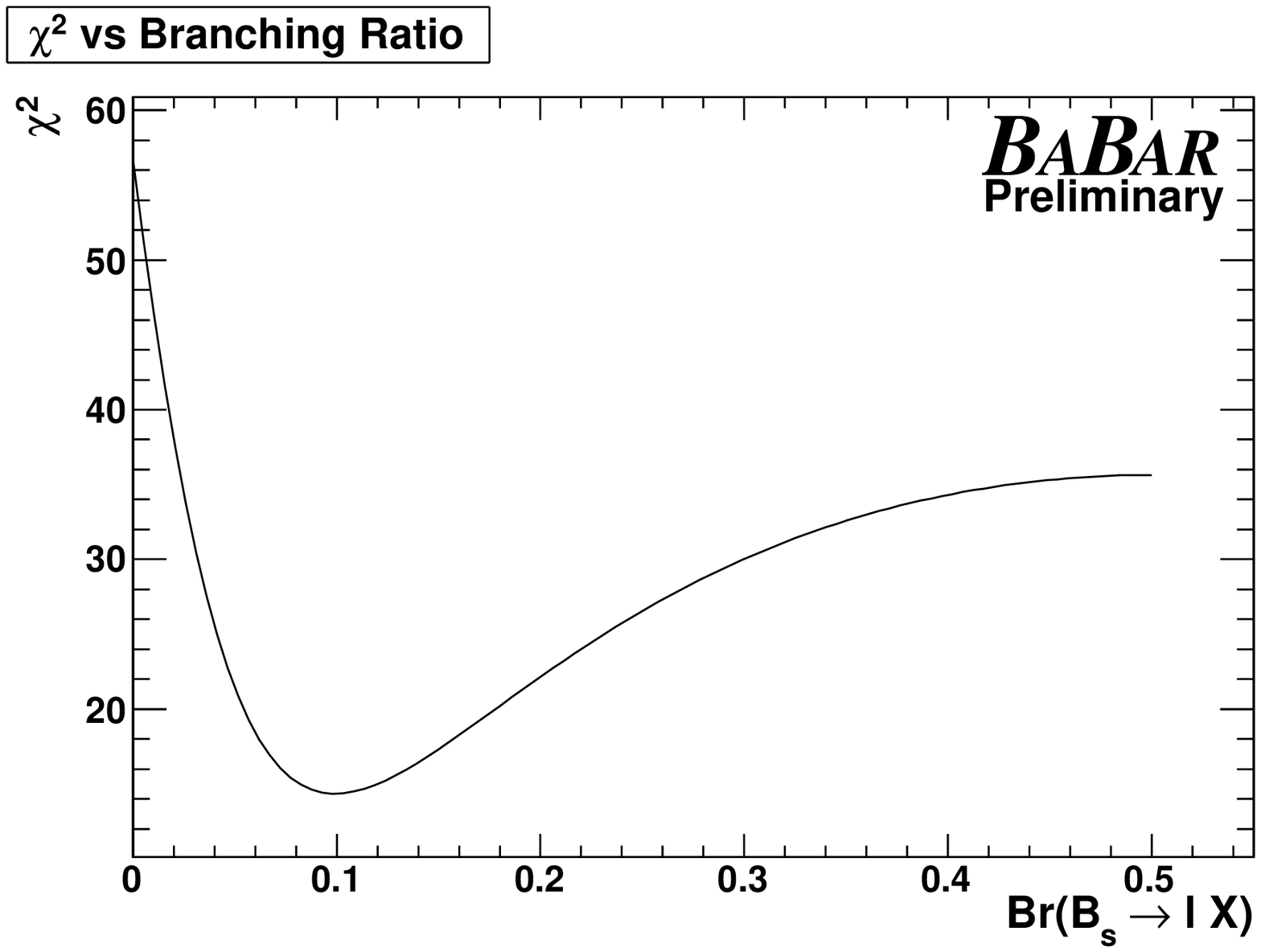}
\caption{The $\chi^2$ as a function of $\Bl$ drawn in the physical region (restricted to less than 0.5 because of the definition of $\Bl$ as the average of the branching fractions to $e$ and $\mu$.}
\label{fig:chi2}
\end{figure}

Using Eqs. \ref{eqn:c1} and \ref{eqn:c2}, $f_s$ is extracted as a function of $E_{\rm CM}$, and is shown in Fig. \ref{fig:fs}. The production fraction is observed to peak in the region of the $\Y5S$ mass, and is small elsewhere. This is in qualitative agreement with the coupled-channel analysis presented in Ref \cite{coupledchannel}, which predicts such a behavior. 

Once $f_s$ is known in each bin, a $\chi^2$ is formed by extracting $\epsilon^s_{\phi\ell}P(B_s\Bbar_s \to \phi\ell X)$ at each point and comparing it to an expected value expressed as a function of the unknown $\Bl$ as well as a dozen other branching fractions which are either obtained from Ref. \cite{PDG} or estimated. The two estimated branching fractions contribute a small amount to the final systematic uncertainty compared to $\mathcal{B}(B_s \to D^{(*)}_s)=(0.93\pm0.25)$ from Ref. \cite{PDG}, which is currently the dominant source of systematic uncertainty. This $\chi^2$ is then minimized with respect to $\Bl$ to obtain a best-fit value of the semileptonic branching fraction. The shape of the $\chi^2$ in the physical region can be seen in Fig. \ref{fig:chi2}. Note that due to the pair of $B_s$ mesons in the event, the quantity $P(B_s \Bbar_s \to \phi X)$ is quadratic in $\Bl$, such that the $\chi^2$ is quartic. This quartic dependence generates the highly asymmetric behavior of the minimum.

The preliminary result obtained in this analysis is  $\Bl = 9.9{ }^{+2.6}_{-2.1}\stat{ }^{+1.3}_{-2.0}\syst$.

\section{Conclusion}

The study of semileptonic decays of $B$-flavored hadrons remains an active area of research in the $B$-factories, now that their respective datasets are complete. In the present work the preliminary results of the study of exclusive $B$ decays to tau and inclusive $B_s$ semileptonic decays has been presented. Though both studies have systematic uncertainties of similar size to their statistical uncertainties, both have clear paths for improvement at the planned super flavor factories. 



\bigskip 

\begin{thebibliography}{9}   

\bibitem{Aubert:2001tu}
B.\ Aubert {\em et al.} (\babar\ Collaboration),
Nucl. Instr. Methods Phys. Res., Sect. A {\bf 479}, 1 (2002).

 \bibitem{oldresult}
 B.~Aubert {\it et al.} (The \babar\ Collaboration),
 Phys.\ Rev.\ D {\bf 79}, 092002 (2009).
 
 \bibitem{RD}
 Nierste, Trine and Westoff,
 Phys.\ Rev.\ D {\bf 78}, 015066 (2008).
 
 \bibitem{tanaka}
 Tanaka and Wanatabe, arXiv:1005:4306[hep-ph].

\bibitem{PDG}
 K.~Nakamura {\it et al.} (Particle Data Group),
 J.\ Phys.\ G {\bf 37}, 075021 (2010).
 
 
 \bibitem{kernel}
 K.S.~Cranmer, Comput. Phys. Commun. {\bf 136}, 198 (2001).
 
 \bibitem{coupledchannel}
N.A.~T\"ornqvist, Phys. Rev. Lett. {\bf 53}, 878 (1984).
 
 

\end{thebibliography}

\end{document}